\documentclass{article}

\usepackage[nonatbib, preprint]{nips_2018}

\usepackage[utf8]{inputenc} % allow utf-8 input
\usepackage[T1]{fontenc}    % use 8-bit T1 fonts
\usepackage{hyperref}       % hyperlinks
\usepackage{url}            % simple URL typesetting
\usepackage{booktabs}       % professional-quality tables
\usepackage{amsfonts}       % blackboard math symbols
\usepackage{nicefrac}       % compact symbols for 1/2, etc.
\usepackage{microtype}      % microtypography
\usepackage{graphicx}
\usepackage{wrapfig}
\usepackage{stmaryrd}
\usepackage{tikz}

\title{Predicting thermoelectric properties from crystal graphs and material descriptors -- first application for functional materials}

\author{
    Leo Laugier \textsuperscript{1} \\
    I2R, A*STAR \\
    Singapore  \\
    \texttt{leojl@i2r.a-star.edu.sg} \\
\And
    Daniil Bash \textsuperscript{1} \\
    IMRE, A*STAR \\
    Singapore  \\
    \texttt{daniil.bash@u.nus.edu.sg} \\
\And
    Jose Recatala \\
    IMRE, A*STAR \\
    Singapore\\
    \texttt{stu-joserg@imre.a-star.edu.sg} \\
\And
    Hong Kuan Ng \\
    IMRE, A*STAR \\
    Singapore\\
    \texttt{stu-nghk@imre.a-star.edu.sg} \\    
\And
    Savitha Ramasamy \\
    I2R, A*STAR \\
    Singapore\\
    \texttt{ramasamysa@i2r.a-star.edu.sg} \\
\And
    Chuan-Sheng Foo \\
    I2R, A*STAR \\
    Singapore\\
    \texttt{foocs@i2r.a-star.edu.sg} \\
\And
    Vijay R Chandrasekhar \textsuperscript{2}\\
    I2R, A*STAR \\
    Singapore \\
    \texttt{vijay@i2r.a-star.edu.sg} \\
\And
    Kedar Hippalgaonkar  \textsuperscript{2}\\
    IMRE, A*STAR \\
    Singapore\\
    \texttt{kedarh@imre.a-star.edu.sg} \\
}

\begin{document}

%\begin{tikzpicture}
%\useasboundingbox (0,0) rectangle (.01cm,.01cm);
%\draw (0,0) -- +(.25\linewidth,0);
%\draw[red] (0,-.75\textheight) node{Hahah};
%\end{tikzpicture}

\maketitle

\begin{abstract}
  We introduce the use of Crystal Graph Convolutional Neural Networks (CGCNN), Fully Connected Neural Networks (FCNN) and XGBoost to predict thermoelectric properties. The dataset for the CGCNN is independent of Density Functional Theory (DFT) and only relies on the crystal and atomic information, while that for the FCNN is based on a rich attribute list mined from Materialsproject.org. The results show that the optimized FCNN is three layer deep and is able to predict the scattering-time independent thermoelectric powerfactor much better than the CGCNN (or XGBoost), suggesting that bonding and density of states descriptors informed from materials science knowledge obtained partially from DFT are vital to predict functional properties.
\end{abstract}

\section{Introduction}
\textcolor{white}{\footnote{equal contributions}}
\textcolor{white}{\footnote{joint corresponing authors}}
The discovery of novel functional materials is an arduous process, which often relies on fortune and serendipity for success \cite{1}. 
In this paradigm, the scientific community is pushing for rational,data-driven accelerated design of materials
\cite{2,3}. Development of density functional theory (DFT) has proven to be a vital step towards high-throughput (HT) discovery of materials \cite{4}.
However, despite its wide acceptance, DFT is computationally intensive and still needs much work before representing experimental values accurately. 
This leads to non-systematic errors, which can be quite substantial \cite{5}.
Nowadays, a variety of machine learning (ML) algorithms provide an alternative to costly and complex DFT calculations, providing similarly accurate results in a fraction of time \cite{6}. 
Machine learning can be a powerful tool for material scientists, since it is capable of unravelling underlying and previously unknown correlations between \emph{a priori} unrelated material descriptors \cite{6}.
Deployment of ML algorithms in materials science is an emerging area of research \cite{7}.
Some of them include the prediction of the stability of crystal structures \cite{8,9} as well as crystal properties like melting points in binary mixtures \cite{10}, vibrational entropies and free energies of crystalline compounds \cite{11}, band gaps of a specific type of materials e.g. perovskites \cite{12}, as well as the discovery of new materials, for instance metallic glasses \cite{13}, lead-free hybrid organic-inorganic perovskites \cite{14} or new molecules for organic flow battery electrolytes \cite{15}.

In this work, we explore a hitherto unexplored idea of using materials descriptors to directly predict functional properties. 
Considering thermoelectrics (TE) as an example, we use Crystal Graph Convolutional Neural Networks (CGCNN) to infer functional thermoelectric property of materials in an attempt to bypass computationally expensive DFT calculations. We compare this approach to using DFT-derived features with a Fully Connected Neural Network (FCNN) or gradient boosting.

\section{Background on Thermoelectrics}

The performance of a TE material is determined by its figure of merit, $ZT = S^2\sigma T / \kappa$, where $S, \sigma, T$, and $\kappa$ are the Seebeck coefficient, electrical conductivity, temperature and thermal conductivity, respectively. 
Traditionally, full Boltzmann Transport Equations (BTE) \cite{16} can be used to predict the Seebeck and electrical conductivity, however these have many inherent assumptions and can also be time-consuming as integration over the full Brillouin Zone is required. 
The input used in these calculations is the material's band-structure. 
A full BTE calculation implemented using DFT calculations as input \cite{17} cannot serve as a comprehensive prediction tool since estimation of relaxation times of charge carriers is computationally expensive and an even harder challenge. 
Hence, using a constant relaxation time approximation (CRTA), typically, DFT band-structure inputs are used along with linearized BTE calculations to predict $S^2\sigma / \tau\textsubscript{0}$ (henceforth called the power factor) where $\tau\textsubscript{0}$ is the constant relaxation time. Note,  this deviates from convention where power factor includes the relaxation time, but we call it such for ease of interpretability.  Accounting for the complexity of the band-structure in the BTE calculations \cite{32} and high-throughput screening to estimate relaxation time and power factor \cite{3} are state-of-the-art, but no ML tools have been used for this purpose yet. 

Therefore, the electronic power factor as defined above is the first screening parameter that links the material's electronic structure to its $ZT$. Leveraging upon detailed calculations performed by Ricci et. al. and using their open-source dataset \cite{18}, we use these theoretically computed power factors as outputs for our two supervised ML approaches, as shown in {\it Figure 1} below. 
Therefore, we use the DFT + BTE power factor predictions as ground truth. Further, we compared the accuracy of the CGCNN to a fully connected neural network (FCNN) algorithm, trained on a richer set of inputs with additional descriptors, obtained from Materials Project Database \cite{2} using Python package ``Matminer'' \cite{3} and data by Ricci, F. et al \cite{18}. 
We find that the accuracy of our predictions of the true electronic power factor are better than recent work on using atomic and bonding descriptors to predict the Seebeck coefficient only (30-46\% for different temperature ranges). \cite{19}

\section{Dataset}

%\subsection{Mining online materials databases}

The open-source Python library named `Matminer' \cite{20,21} provides a unified API, and acts as an intermediary between the user and four commonly used open-source materials databases: Citrination,\cite{22} Materials Project (MP),\cite{23} Materials Data Facility (MDF),\cite{24} and Materials Platform for Data Science (MPDS).\cite{25} Using Matminer and Ricci \textit{et. al.}'s dataset \cite{18} with filtering and cleaning, the final database comprising 7230 indexed compounds with 28 features that include atomic, structure and bonding descriptors was generated. In addition, by accounting for the doping type, doping level, temperature and crystal direction for each compound as input features, 2,819,682 datapoints were created. 
%Particularly relevant is Matminer's data featurization, capable of transforming materials data into desired feature descriptors (e.g. composition, crystal structural, density of electronic states, band structure, or atomic sites). 
\begin{figure}
\centering
\includegraphics[width=0.7\linewidth]{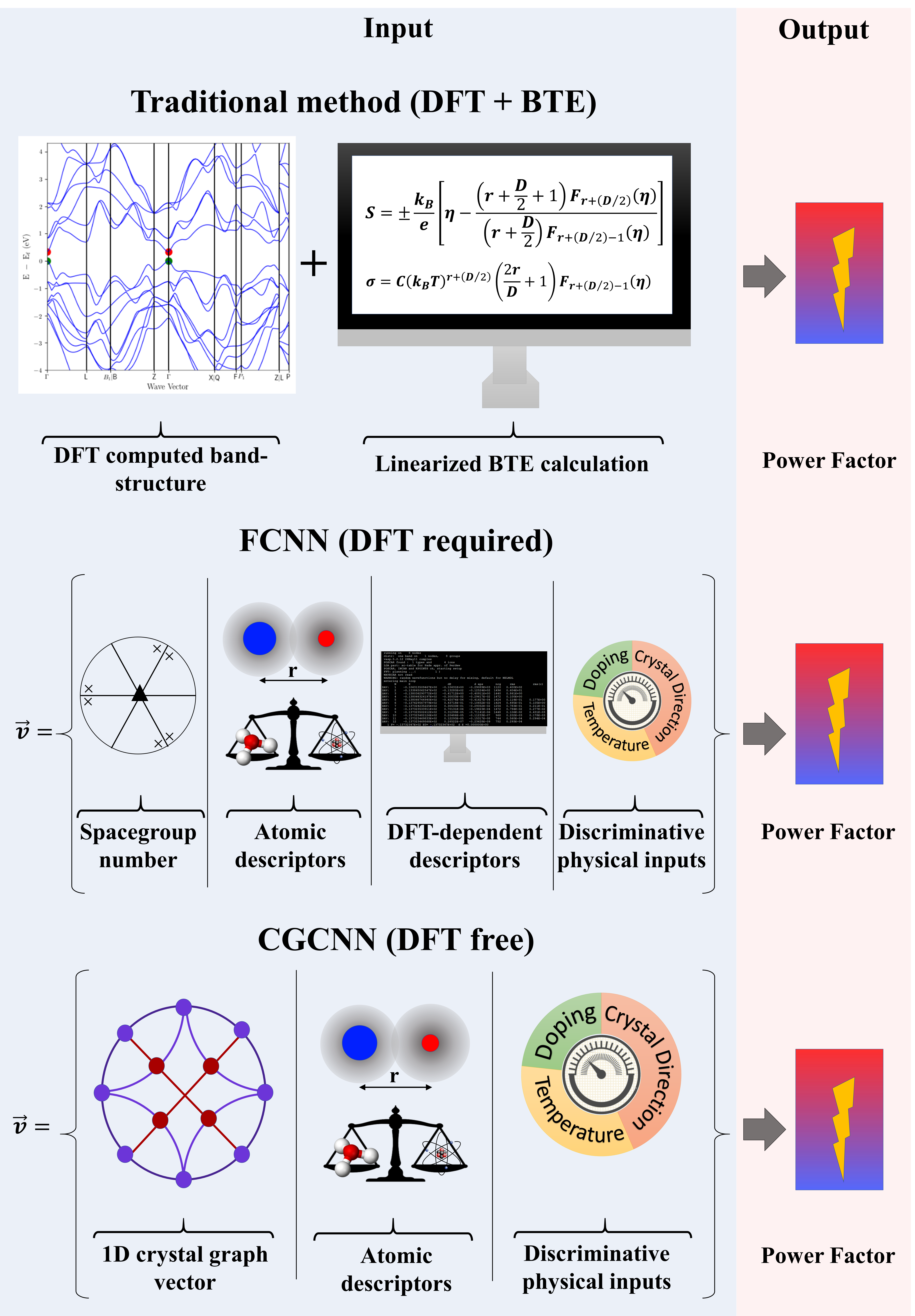}
\caption{\label{fig:features}Graphical representation of the one-dimensional vector used for ML predictions of thermoelectric power factor.}
\end{figure}

\subsection{Workflow for dataset generation}

Traditionally, the calculation of the electronic power factor for a material is obtained in two steps. First, the electronic band structure properties (e.g. effective mass, reduced fermi energy, valley degeneracy) are calculated from DFT. 
Second, the TE properties (Seebeck coefficient and electrical conductivity) are calculated using the linearized Boltzmann Transport Equations (BTE) such as self-programmed scripts or BoltzTraP.\cite{17,26,27}
Recently, models have exploited external databases (such as Materials Project)\cite{23} as well as BTE packages (such as BoltzTraP2)\cite{17,27} for the mining of required DFT parameters and calculation of TE properties. 
This combination of DFT and BTE has allowed calculations of TE properties, yielding the largest computational database of electronic transport properties, to the best of our knowledge.\cite{18}

Our initial dataset was obtained from the work of Ricci \emph{et. al}\cite{18} and the general workflow is depicted in {\it Figure 2}. 
It contains around 23,000 .json files, each comprising multi-level data from the Materials Project Database (MPD). 
Each of these files corresponds to a specific material. 
The full description of the original data can be found in the original work.\cite{18}

\begin{figure}[!htbp]
\centering
\includegraphics[width=\linewidth]{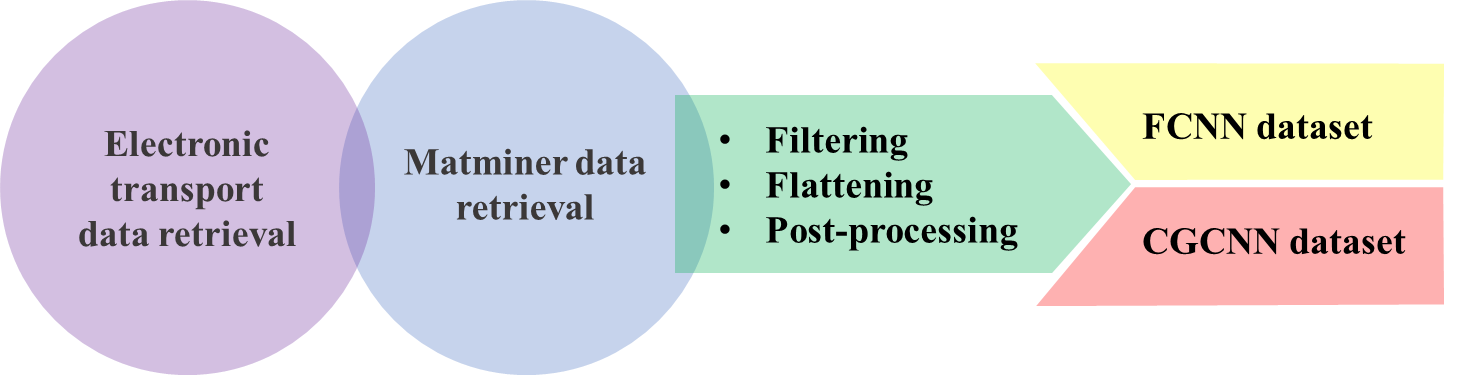}
\caption{\label{fig:workflow}Schematic representation of the general workflow utilised for the construction of the ML datasets.}
\setlength{\belowcaptionskip}{-10pt}
\end{figure}

For our study, the following list of key parameters was extracted: cond\_doping, seebeck\_doping, mp\_id, pretty\_formula, cif\_structure, spacegroup.number, volume, nsites, band gap, GGA/GGA+U. 
Cond\_doping and seebeck\_doping represent the electrical conductivity (divided by the constant relaxation time, $\sigma/\tau_0$) and the Seebeck $S$ comprised of the following nested dictionaries: electron or hole doped (n/p), temperature (100 - 1300K) and doping concentrations (1x10\textsuperscript{16-20} cm\textsuperscript{-3}) and crystal direction (x,y,z). 
Overall, each compound contains 390 data points for cond\_doping and seebeck\_doping, each representing different discriminative physical inputs(n/p, T and x/y/z).  The extracted data is flattened to unravel all nested parameters into a single dimensional vector. The full list of descriptors used is shown in {\it Figure 3}.

\subsection{Filtering}

The dataset was then filtered to only represent stable semiconductor materials. The value of band gap was set to be within 0.16-4 eV. This interval was chosen because good TE performance is found mostly in semiconductors with band gap in this range. This range for band gap was corrected by a factor of 1.6, taking into account the errors from DFT calculations. Also, the value of e\_above\_hull was limited to be less than 0.05 eV/atom, thus representing only stable compounds.  In addition, compounds with no data on DOS or efermi were filtered out, arriving at the final dataset comprised of 7230 indexed compounds. 
Next, an output column labelled power factor was calculated by considering the product of the seebeck\_doping with square of the cond\_doping ($S^2 \sigma / \tau_0$). 
The values of spacegroup number, crystal direction and direct/indirect bandgap were one-hot encoded (OHE). Here, the dataset was further split into two subsets: one for the FCNN and the other for CGCNN.

For the FCNN approach, the cif data was removed from the dataset and replaced by the OHE representation of the spacegroup, while in the CGCNN approach all descriptors derived from DFT were deleted, including the spacegroup number. 
The final list of descriptors can be found in {\it Figure 3}.  Hence, the final datasets were transformed to include the doping level and type, temperature and crystal direction as inputs for each material in consideration, yielding nearly 3 million rows of unique combinations of values .

\begin{figure}[!htbp]
\centering
\includegraphics[width=\linewidth]{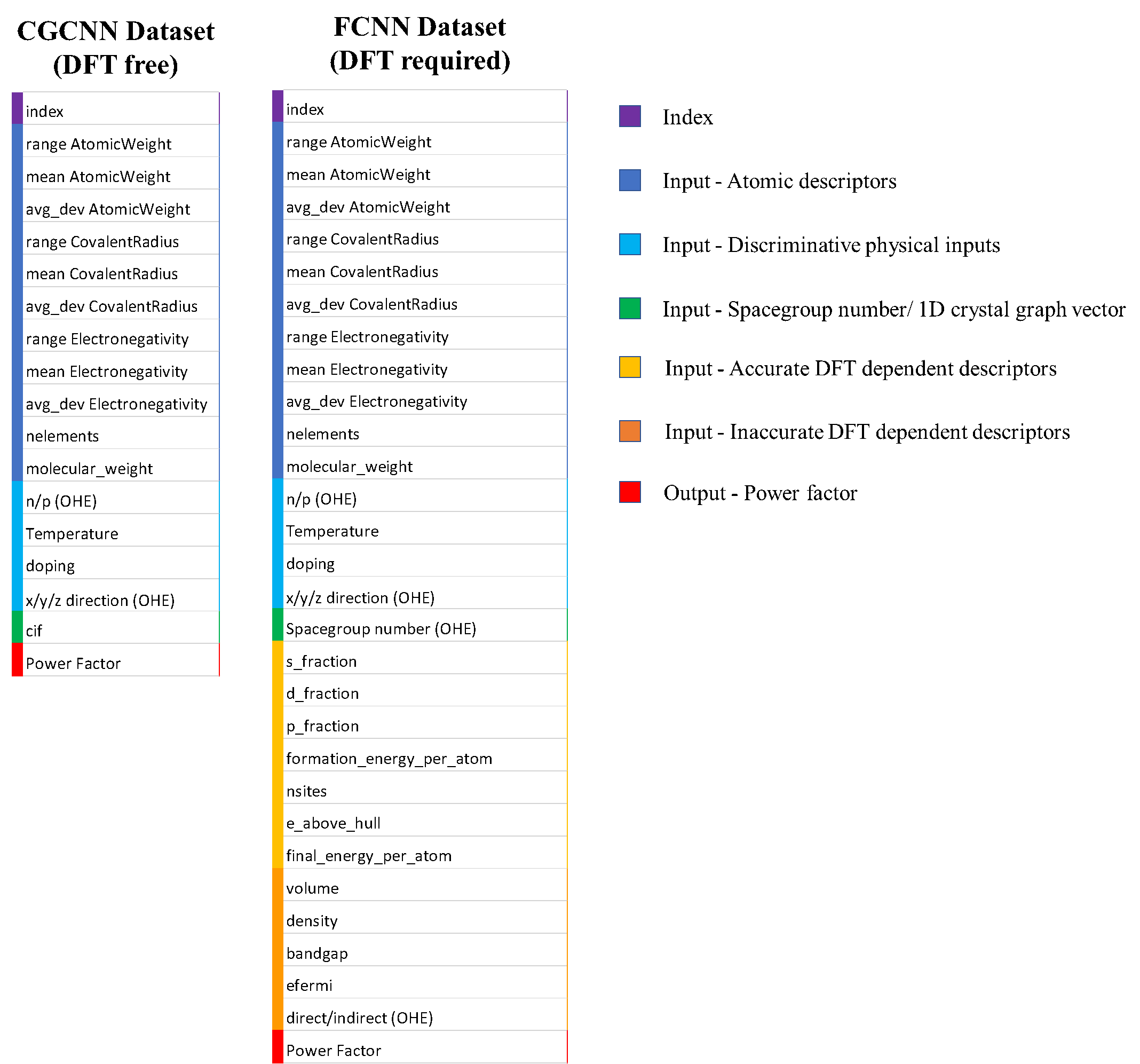}
\caption{\label{fig:datasets}Full list of the descriptors, used in training datasets for the different ML algorithms.}
\end{figure}

\section {Methods}

\subsection{Inputs and Methods (Fully Connected NN vs CGCNN)}
\subsubsection{Data preprocessing}
We have two kinds of features that we preprocess differently.

First, graphs are described by atoms and bonds between them.
The atomic and bonding data is encoded in $v^{l=0}_{i}$ and $u_{{(i,j)}_k}$, which represent nodes of the crystal and its edges respectively. 
The node feature vector ($v_0$) comprises one hot encoded elemental properties for each atom in the crystal unit cell, such as the position of the atom in the periodic table, its size and chemical character. The bonding between two atoms i and j is described by means of the bond feature vector ($u_{{(i,j)}_k}$). The full description of the initialization of vertices and edges can be found in the supplementary table II (atoms) and table III (bonds) of \cite{28}.

Second, we use additional feature descriptors that are standardized for XGBoost and CGCNN with a standard scalar. The features for the FCNN were either scaled with a specific scalar or not scaled depending on the hyperparameter setting (see 5.1.3).

\subsubsection{Models}
As a strong, non neural-network baseline, we used XGBoost \cite{31}, a popular implementation of gradient boosting that has achieved high predictive performance across a range of tasks. 

Graph CNN have shown to be useful to solve fundamental learning problems such as graph clustering and sub-graph matching \cite{29}. The advantage of this architecture is to learn a vector representation of variable-length graphs by encoding, at each layer and for each vertex, information coming from the one hop neighbors. After $L$ layers, the vertices have received the information coming from their L hop neighborhood which allows to build a global representation of the graph by pooling the last updated representation of nodes. Xie \emph{et al.} \cite{28} have developed their specific Crystal Graph CNN architecture for the prediction of material properties, that we took over for the prediction of functional properties of compounds. 

We compared the relatively novel CGCNN with more traditional Machine Learning and Deep Learning models that are XGBoost and the fully connected neural network (FCNN).

As the distribution of the logarithm of the power factor (PF) targets seems to follow a Gaussian distribution, we predicted log(PF) rather than PF. For the XGBoost and CGCNN we standardized log(PF).

\subsection{Architecture of the CGCNN}
Each convolutional layer $(l)$ of CGCNN is composed of two weights matrices ($W_{f}^{l}$ and $W_{s}^{l}$) and two bias vectors ($b_{f}^{l}$ and $b_{s}^{l}$) that are learned during training. The convolutional computation follows equation \ref{eq:1}:

\begin{equation} \label{eq:1}
v_{i}^{l+1} = v_{i}^{l} + \sum_{(j,k) \in neighborhood(i) \times \llbracket 1 ; K \rrbracket} \sigma(z^{l}_{(i,j)_{k}} W_{f}^{l} + b_{f}^{l}) \odot g(z^{l}_{(i,j)_{k}} W_{f}^{l} + b_{s}^{l})
\end{equation}

where $K$ is the maximal number of bonds connecting two atoms, $g$ is the softplus activation function, $\sigma$ is the sigmoid activation function, $v_{i}^{l}$ is the representation of node $i$ at layer $l$ and $z^{l}_{(i,j)_{k}}$ is the concatenation of $v_{i}^{l}$, $v_{j}^{l}$ and $u_{(i,j)_k}$.

At the end of the last convolutional layer, we pool the last representation of the vertices $(v_{i}^{L})_{i}$ to get a single vector representing the graph. We append to this vector the other descriptors of the molecule (see Figure 3). Eventually this vector is input to fully connected layers with softplus activation functions to predict log(PF). The loss function is the Mean Squared Error between prediction and the logarithm of ground truth log(PF).

\subsection{Hyperparameter Tuning}
%\textbf{Add a few sentences briefly describing hyperband and what it does.}

%\textbf{For hyperparameters and best ones put in a single table with 3 columns: Hyperparameter | Search range | Best value}

We used the Hyperband \cite{30} hyperparameter tuning strategy for XGBoost and FCNN and we did a random search for tuning the hyperparameters of CGCNN.

Hyperband is a bandit-based approach to hyperparameter tuning, that randomly samples the parameter space. It is based on the following principles: it first starts to run configurations on few iterations and uses this information to keep only hyperparameter settings that will give the best performance when running on more iterations. Hyperband is more efficient than a purely random search but also faster than a grid search.

\section{Experiments}
\subsection{Experimental setup}
The evaluation metric used for selecting the best set of hyperparameters and comparing the three models is the relative Mean Absolute Error (MAE) between the exponent of the prediction and the  ground truth PF. 

\begin{equation} \label{eq:2}
relMAE((\vec{y_{i}})_{_{i}},(\vec{\hat{y}_{i}})_{_{i}}) = \frac{1}{n} \times \sum_{i} \mid \frac{\vec{y_{i}} - \vec{\hat{y}_{i}}}{\vec{y_{i}}} \mid
\end{equation}

where $\frac{\vec{x_{1}}}{\vec{x_{2}}}$ denotes here the division element-wise of $\vec{x_{1}}$ vector by $\vec{x_{2}}$, $(\vec{y_{i}})_{_{i}}$ are the ground truth targets and $(\vec{\hat{y}_{i}})_{_{i}}$ are the corresponding predictions.

We use a relative error because, as PF is in the order of $10^{22}$ and we train the models on log(PF). An absolute MAE of $10^{20}$ corresponds to a relative MAE of 1\%. 

We split the dataset into training, validation and testing sets. 60\% of the data (1691810 samples) were used for training, 20\% (563936 samples) were used for validation and 20\% (563936 samples) were in the testing set. We used the same split for the three models.

During hyperparameter optimization, for each model we select the architecture with the lower relative MAE on the validation set and eventually we compare the performance of the three models, under the selected architecture, on the test set. 

Table \ref{table:hpcgcnn}, \ref{table:hpxgboost}, and \ref{table:hpfcnn} summarize the hyperparameter tuning respectively on CGCNN, XGBoost and FCNN. 

\begin{figure}[!htbp]
\centering
\includegraphics[width=\linewidth]{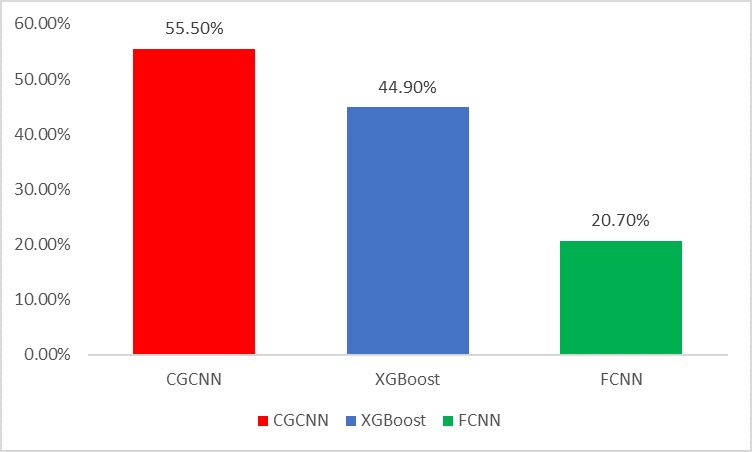}
\caption{\label{fig:results} Comparison of the relative MAE for the different models.}
\end{figure}

Figure \ref{fig:results} shows the evaluation of the three models on the test set. 

%subsubsection{CGCNN}
\begin{table}
    \centering
    \begin{tabular}{|p{4cm}|l|l|}
        \hline
        \textbf{Hyperparameter} & \textbf{Range} & \textbf{Best} \\
        \hline
        Number of convolutional layers & 1--5 & 2 \\
        \hline
        Number of fully connected layers & 1--5 & 1  \\
        \hline
        Minibatch size & $\{128 , 256 \} $ & 256  \\
        \hline
        Number of hidden atom features in convolutional layers & $\{8 , 16 , 32 , 64 , 128 , 256 \} $ & 256  \\
        \hline
        Number of hidden features after pooling & $\{8 , 16 , 32 , 64 , 128 , 256 \} $ & 128  \\
        \hline
        Weight decay & $ \left[ 0 , 1 \right] $ & 0.97  \\
        \hline
        Dropout & $ \left[ 0 , 1 \right] $ & 0.366  \\
        \hline
        Momentum & $ \left[ 0 , 1 \right] $ & 0.491  \\
        \hline
        Optimizer & SGD or ADAM & ADAM  \\
        \hline
    \end{tabular}
    \\[10pt]
    \caption{Hyperparameters for CGCNN.}
    \label{table:hpcgcnn}
\end{table}

%The hyperparameters for CGCNN were the number of convolutional layers in $\llbracket 1 ; 5 \rrbracket $, the number of fully connected layers in $\llbracket 1 ; 5 \rrbracket $, the mini batch size in $\{128 ; 256 \} $, the number of hidden atom features in the convolutional layers and the number of hidden features after pooling in $\{8 ; 16 ; 32 ; 64 ; 128 ; 256 \} $, the weight decay, the dropout and the momentum in $ \left[ 0 ; 1 \right] $ and the optimizer being either Stochastic Gradient Descent or ADAM.

%\subsubsection{XGBoost}
\begin{table}
    \centering
    \begin{tabular}{|p{7cm}|l|l|}
        \hline
        \textbf{Hyperparameter} & \textbf{Range} & \textbf{Best} \\
        \hline
        maximum tree depth & 2---10 & 6 \\
        \hline
        minimum sum of instance weight (hessian) needed in a child & 1--10  & 5  \\
        \hline
        learning rate & $\left[ 0.01 , 0.2  \right]$ & 0.17  \\
        \hline
        minimum loss reduction required to make a further partition on a leaf node of the tree & $\left[ 0 , 1 \right]$ & 0  \\
        \hline
        L1 regularization term on weights & $\left[ 10^{-10} , 1 \right]$ & 0  \\
        \hline
         L2 regularization term on weights & $\left[ 0.1 , 10 \right]$ & 6.87  \\
        \hline
        initial prediction score of all instances & $\left[ 0.1 , 0.9 \right]$ & 0.5  \\
        \hline
        balancing of positive and negative weights & $\left[ 0.1 , 10 \right]$ & 1  \\
        \hline
        subsample ratio of the training instance & $\left[ 0.5 , 1 \right]$ & 1  \\
        \hline
        subsample ratio of columns for each split, in each level & $\left[ 0.5 , 1 \right]$ & 0.93  \\
        \hline
    \end{tabular}
    \\[10pt]
    \caption{Hyperparameters for XGBoost.}
    \label{table:hpxgboost}
\end{table}

%The hyperparameters for XGBoost are the maximum tree depth in $\llbracket 2 ; 10 \rrbracket$, the minimum sum of instance weight (hessian) needed in a child in $\llbracket 1 ; 10 \rrbracket$, the learning rate in $\left[ 0.01 ; 0.2  \right]$, the minimum loss reduction required to make a further partition on a leaf node of the tree in $\left[ 0 ; 1 \right]$, the L1 regularization term on weights in $\left[ 10^{-10} ; 1 \right]$, the L2 regularization term on weights in $\left[ 0.1 ; 10 \right]$, the initial prediction score of all instances in $\left[ 0.1 ; 0.9 \right]$, the balancing of positive and negative weights in $\left[ 0.1 ; 10 \right]$, the subsample ratio of the training instance and the subsample ratio of columns when constructing each tree and the subsample ratio of columns for each split, in each level in $\left[ 0.5 ; 1 \right]$.

%\subsubsection{FCNN}

\begin{table}
    \centering
    \begin{tabular}{|p{4cm}|p{4cm}|l|}
        \hline
        \textbf{Hyperparameter} & \textbf{Range} & \textbf{Best} \\
        \hline
        shuffle the training data before each epoch or not & True or False & False \\
        \hline
       number of fully connected layers & 1---5 & 3  \\
        \hline
        size of layers & 1---100 & 50 (l1), 41 (l2), 41 (l3)  \\
        \hline
        activation of layer & ReLU, tanh or sigmoid & ReLU (l1), tanh (l2), sigmoid (l3)  \\
        \hline
        mini batch size & $\{16, 32, 64, 128, 256\}$ & 256 \\
        \hline
        L1 regularization term on weights & $\left[ 10^{-10} , 1 \right]$ & 0  \\
        \hline
         L2 regularization term on weights & $\left[ 0.1 , 10 \right]$ & 6.87  \\
        \hline
        initial prediction score of all instances & $\left[ 0.1 , 0.9 \right]$ & 0.5  \\
        \hline
        initialization of layers & uniform, normal, glorot uniform, glorot normal, He uniform or He normal & glorot normal \\
        \hline
        loss function & MSE or MAE & MSE  \\
        \hline
        optimizer & RMSProp, ADAGRAD, ADADELTA, ADAM or ADAMAX & RMSProp  \\
        \hline
        scaler & None, standard, MinMax, MaxAbs or Robust & Robust  \\
        \hline
    \end{tabular}
    \\[10pt]
    \caption{Hyperparameters for FCNN.}
    \label{table:hpfcnn}
\end{table}
%The hyperparameters for FCNN are whether to shuffle the training data before each epoch or not, the number of fully connected layers between in $\llbracket 1 ; 5 \rrbracket$, the sizes of each layer in $\llbracket 1 ; 100 \rrbracket$, the mini batch size in $\{16, 32, 64, 128, 256\}$, the activation of each layer being a RELu, a sigmoid or a hyperbolic tangent, the initialization of the dense layers being either uniform, normal, glorot uniform, glorot normal, He uniform or He normal, the loss function being either the mean absolute error or the mean squared error, the optimizer being either RMSProp,  ADAGRAD, ADADELTA, ADAM or ADAMAX, the scaler of the inputs being either absent, a standard scaler, a MinMax scaler (scaling between 0 and 1), a MaxAbs scaler (scaling each feature by its maximum absolute value) or a Robust scaler scaling features using statistics that are robust to outliers.

%\subsection{Predictive Performance and Results}
%\textbf{Replace by a single bar chart with one bar for each method; best hyperparameters should be in the tables}

\section{Discussion}

The results show that we should opt for FCNN when predicting DFT+BTE thermoelectric power factor. It seems that the CGCNN operating on simple DFT-free descriptors of the compound cannot bypass DFT calculations since CGCNN performed worse than both FCNN and XGBoost. FCNN is better than XGBoost but takes time to converge, howewer has the highest prediction capability.

Furthermore, the current dataset has some limitations.
It includes duplicates of the same compound, but with different crystal structures, some of which might not be stable under experimental conditions, while the energy above the hull parameter mitigates this problem somewhat. 
Also, some compounds in the dataset do not exactly match the selection criteria in real life, while when computed theoretically they have unrealistic values (like semiconducting form of N\textsubscript{2}, or S). 
As previously discussed, the limitations of DFT calculations lead to non-systematic errors. 
Moreover, BTE calculations assume CRTA, where all scattering events that can influence electron conduction such as impurity scattering, phonon scattering, etc are included in this parameter.\cite{18} The CRTA indicates that BTE is modelled to be isotropic and energy-independent, and this resulted in the key limitation, where calculated TE properties are far from experimental values for a number of materials. Extending the simulation capabilities to produce more accurate relaxation times is an ongoing exciting field of research, where ML tools can contribute and also benefit from.  Experimental datasets such as Citrination \cite{22}  are still developing, and are limited in the range of compounds, their doping levels and temperature.    

\section{Conclusion and future work}
We've shown that machine learning tools can predict the electronic power factor of thermoelectric materials based on theoretically calculated datasets. It is apparent that models based on simple atomic and bonding information without detailed DFT inputs cannot predict TE properties well. This is the first such work on thermoelectrics and can be extended to other functional materials provided a dataset is available.  

In future work more detailed filtering criteria as well as additional featurizers and materials descriptors will be explored.  The next step constitutes feature importance targeting discovery of which descriptors of the compound has most influence on the power factor, informing predictive design that experimentalists can access to find better thermoelectric compounds. Training deep learning models on more epochs and continuing the hyperparameter search would find more accurate models.

Evaluating other Graph CNNs (such as Graph CNN updating edge representation at each layer) could help discover more suitable architectures for the specific application of predicting PF or other thermoelectrical and functional material properties. Future work will involve conducting a thorough study on experimental data. This would allow us to compare simulated values of power factor with  experimental observations and then, once a large enough database of empirical power factor will have been generated, we may train machine learning and deep learning models on this actual ground-truth database towards accurate predictions of experimental thermoelectric properties.

\end{document}